# TIEboard: A Digital Educational Tool for Kids Geometric Learning

AROOJ ZAIDI, Keio Graduate School of Media Design, Japan
GIULIA BARBARESCHI, Keio Graduate School of Media Design, Japan
KAI KUNZE, Keio Graduate School of Media Design, Japan
YUN SUEN PAI, School of Computer Science, University of Auckland, New Zealand
JUNICHI YAMAOKA, Keio Graduate School of Media Design, Japan

Tangible User Interfaces have shown potential in supporting the acquisition of key concepts in computing and mathematics while fostering engagement in young learners, but these approaches are less commonly utilised in the context of geometry. In this paper we introduce TIEboard, an interactive device to promote early learning of basic geometry concepts. TIEboard draws inspiration from traditional geoboards and lacing toys to leverage children's familiarity with these traditional tools. It employs instructional lights to guide children in creating shapes using colourful threads of optical fiber. The use of conductive materials allows the system to detect lacing activity and provide feedback in real-time. TIEboard incorporates six interaction modes of varying difficulty based on an incremental learning framework. The study evaluated TIEboard's effectiveness in supporting early geometric learning, facilitating creativity and promoting collaboration among 16 children aged 5-9.

CCS Concepts: • **Human-centered computing** → **User interface design**; • **Applied computing** → *Interactive learning environments*; • **Hardware** → PCB design and layout.

Additional Key Words and Phrases: Tangible User Interface(TUI), Children Education, Interaction Design, Geometry Learning

## 1 INTRODUCTION

Learning mathematics is not simply about mastering numbers, but encompasses many other competencies which are essential for young children to understand the world [52]. Geometry, as emphasized by the National Council of Teachers of Mathematics (NCTM) [21], serves as a platform for recognizing, understanding, and classifying geometric objects, along with comprehending relationships and visualizing shapes. Both Van Hiele and Piaget advocate for tailoring geometric instructions to align with developmental levels [54, 56, 80]. Despite its significance, geometry is often overlooked in early childhood curricula, with limited engagement strategies and a focus on static concepts due to teachers' emphasis on numeracy and a lack of confidence in the subject [12, 22, 33]. Manipulatives, such as geoboards and lacing toys, help represent abstract mathematical ideas and can enhance children's learning more effectively than mere computer simulations [47, 50, 61, 67]. However,

Authors' addresses: Arooj Zaidi, arooj.zaidi@keio.jp, Keio Graduate School of Media Design, Japan; Giulia Barbareschi, barbareschi@kmd.keio.ac.jp, Keio Graduate School of Media Design, Japan; Kai Kunze, kai@kmd.keio.ac.jp, Keio Graduate School of Media Design, Japan; Yun Suen Pai, yun.suen.pai@auckland.ac.nz, School of Computer Science, University of Auckland, New Zealand; Junichi Yamaoka, Keio Graduate School of Media Design, Japan, yamaoka@kmd.keio.ac.jp.

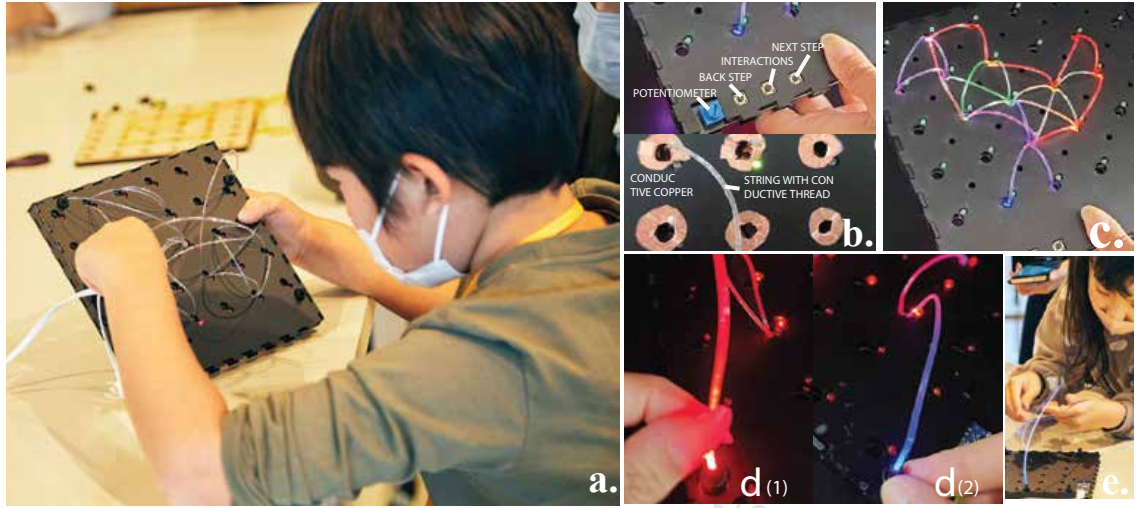

Fig. 1. (a) A child interacting with TIEboard during the workshop to learn the geometric concepts (b) Different buttons to operate TIEboard and sensing connection (c) Intricate shape with TIEboard glowing string (d) 1: Lacing in red color LED to glow string, 2: mixing colors by controlling LED with potentiometer(e) Child interacting with TIEboard glowing string.

while these tools are flexible, they lack embedded scaffolding [43] and guidance[44], hindering the progressive acquisition of geometric concepts and limiting feedback and connections to other representations [44, 72].

Tangible user interfaces (TUIs), also known as digital manipulatives, present a novel paradigm of Human-Computer Interaction (HCI) which integrates computing power and ubiquity in tangible objects [16, 20, 83]. Many researchers have begun exploring the use of digital manipulatives in education, as TUIs offer a way to bring together the benefits of physical manipulation and innovative ways of interaction, thereby enhancing the learning experience and promoting engagement[18, 23, 60, 70]. While TUIs have been examined in education, most tend to concentrate on computing rather than geometry learning, especially for early childhood. Traditional manipulatives, such as geoboards, provide useful tactile engagement but often fall short in offering structured guidance for students to independently advance their understanding of geometric concepts as shown in the Figure 2. TIEboard aims to address this issue by providing a personalized approach that combines tactile and digital feedback to effectively support early geometric learning in both classroom and out-of-school environments (See Figure 2) and Table 1.The device we created is called TIEboard, a digital manipulative which seamlessly integrates computational abilities into a traditional geoboard, designed to promote understanding of abstract geometric concepts [32, 75] amongst children aged 5 to 9 years old. TIEboard is primarily designed for informal learning environments, such as after-school programs, where it can provide flexible and exploratory learning experiences in geometry. Although the system is optimized for these settings, its structured design also makes it adaptable for classroom use, supporting both individual and collaborative learning activities. The device effectively promotes independent learning by integrating built-in scaffolding and real-time feedback, thereby minimizing the need for direct instructor intervention. TIEboard's lacing mechanism combines physical activity with cognitive learning, guiding children as they lace through points that represent key geometric elements (vertices, edges, or shapes). The colored fiber serves as a valuable tool for identifying unusual shapes in specific modes or for breaking down complex figures into simpler components. Upon the successful completion of a shape, the glowing fiber provides motivational visual feedback by lighting up, reinforcing a sense of achievement and encouraging continued engagement. Furthermore, the glowing fiber offers a unique color-mixing experience that fosters playful exploration, thereby enriching interaction with geometric concepts. TIEboard accommodates different levels of skills thanks to its adjustable difficulty, ensuring that children are consistently challenged while not becoming overwhelmed during the learning process [3]. This paper investigates how TIEboard supports children's understanding of geometric concepts, focusing on its ability to provide scaffolded guidance and real-time feedback to teach abstract ideas. Specifically our aim is to address these three research questions: (1) How do children interact with TIEboard to complete basic geometrical tasks? (2) How do children develop and utilize collaboration skills when engaging with TIEboard's activities? (3) How does the free-play mode of TIEboard promote creative learning? To assess TIEboard's effectiveness, we held two workshops with a total of 16 children aged 5 to 9 years. The findings revealed that TIEboard effectively promoted children's understanding of geometric concepts through interactive and engaging exercises. The immediate visual feedback provided by LED assisted children in accurately lace shapes, eventually learning to do so with less guidance once they reached higher levels. Additionally, the customisable thread colors and different interaction modes encouraged innovation and sustained high levels of interest. Collaborative activities encouraged the development of teamwork and communication skills. Overall, the research supported TIEboard's design elements and demonstrated its potential to promote geometric learning. Specifically, this paper makes the following contributions:

(1) The conceptualization and design of TIEboard, a digital educational tool to promote understanding of basic geometric concepts via novel interactions which can be tailored to children's skills
(2) This study relies on research about effective Tangible User Interface (TUI) design, showing how combining digital and physical elements can support organized and step-by-step learning.
(3) The evaluation of TIEboard shows that it effectively enhances children's engagement with geometry through structured activities and free play. This highlights its role as both an instructional tool and a versatile resource for exploring geometric concepts independently, outside traditional classrooms.

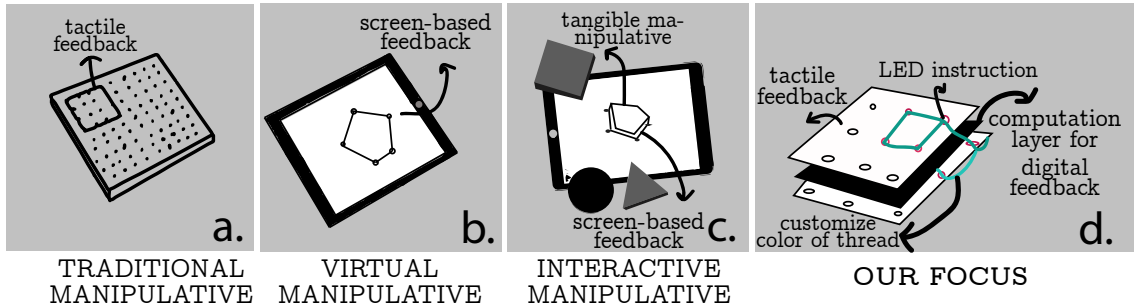

Fig. 2. (a) Traditional interaction (b) Virtual manipulative only using screen for learning (c) Interactive manipulative includes both the screen and tangible manipulative for learning (d) Our focus is to seamlessly combine digital and physical world that functions without screen.

Table 1. Comparison based on the design guidelines presented in [86]

| Design Principle | Traditional Geoboards | Virtual Manipulatives | TIEboard |
|---|---|---|---|
| Generic Structures vs. Real-World Objects | Focus on physical frameworks. | Focus is on digital simulations. | Focuses on early abstract geometric concepts. |
| Level of Abstraction | Low level of abstraction, concrete shapes. | High level of abstraction, digital simulations. | High degree of abstraction, allowing children to create their own analogies. |
| Semantic Association | Overlays or cards featuring recognizable shapes. | Not applicable. | Lacing patterns are rigorously associated with geometric notions. |
| Encourage Analogies | Not applicable. | Limited to software features. | Provides methods to connect lacing patterns with real-life analogies. |
| Modularity | Limited modularity, fixed materials. | Software-dependent modularity. | Modular design with customisable and flexible components. |
| Multi-Sensory Representations | Primarily tactile. | Primarily visual and auditory. | Combines tactile and visual feedback. |
| Coincide I/O | Physical manipulation only. | Digital manipulation separate from physical input. | Manipulation and feedback occur in the same physical space. |
| Synchronous I/O | Sequential interaction. | Delayed feedback based on user input. | Real-time feedback |

## 2 RELATED WORK
### 2.1 Tangible Interfaces in Education

Tangible user interfaces (TUIs) are interactive systems that leverage physical interactions to create engaging experiences through physical manipulation aligned with educational objectives [23, 31] An example is Tern, which enables users to build physical computer programs using interlocking wooden pieces that represent robotic actions [28]. Research such as Michael Eisenberg's Hypergami demonstrates how hands-on interaction with educational tools enhances spatial reasoning, similar to TIEboard's interactive lacing feature aimed at fostering spatial awareness in young learners [19, 39]. Moreover, TUIs that support children's critical thinking rather than provide answers significantly enrich the educational experience [10, 24, 46, 61]. Maria Bartolini Bussi's work on embodied learning in geometry highlights the importance of manipulatives in understanding spatial relationships, reinforcing our hands-on approach in TIEboard [73]. Additionally, Istenic Starcic et al. found that TUIs enhance accessibility and engagement for diverse learning styles, a goal that TIEboard aspires to achieve [74].

Physical materials stimulate both perception and cognition, with TUIs offering multisensory interaction that improves spatial skills and facilitates deeper knowledge acquisition, particularly in geometry [5, 38, 73, 81]. Additionally, TUIs promote collaboration [40], fostering interactive learning environments and enhancing student participation and peer activity visibility [2, 20, 45, 74]. In educational contexts, TUIs adopt various strategies. Some, like SmartStep and FloorMath, connect physical input to computer-based output for fundamental math education [69]. Others, such as TICLE, link physical manipulatives with visualizations on screens to enhance collaborative learning [29, 68]. Digital manipulatives allow direct interaction with devices; for example, SystemBlocks and Flowblocks explore dynamic systems, while Topobo facilitates building programmable robotic creatures [59, 86]. Furthermore, AR-based TUIs, such as Sketched Reality [37], integrate augmented reality sketching with actuated TUIs for interactive drawing. Such hybrid approach offer promising learning opportunities, however, the lack of direct tactile engagement with the projected content can limit exploration. In contrast, TIEboard enables direct tactile interaction through its lacing system, which supports embodied learning. While, the feature of customizing thread colors dynamically adds interactivity and visual engagement without the need for external devices or projection systems, making it more accessible and suitable for early education environments.

Although early education TUIs primarily focus on computing and math, research on geometric learning using TUIs is limited, particularly for children under ten [36, 65].To enhance usage in classrooms and homes, TUIs must facilitate collaboration; provide scaffolding; increase engagement with simple interface; function without personal computers; be in line with the curriculum; support autonomous activities with teacher supervision when necessary; be feasible in the physical space of classroom and stimulate reasoning and be reasonably priced [20]. To date, TUI applications in geometry for learners below the age of 10 years that adhere to these principles are limited. On the other hand, traditional manipulatives are still widely used. The following section will explore traditional manipulatives and their roles in foundational geometric learning.

## 2.2 Traditional Manipulatives in Education

The NCTM [21] recommends the use of manipulatives in the classroom for both geometry and mathematics, highlighting how its efficacy is supported by both learning theory and educational research. The theories of Piaget and Vygotsky [48, 56] suggest that children's cognitive development, especially with abstract concepts like geometry, progresses gradually. This aligns with Areti Panaoura's research, which highlights the need for visual or physical manipulatives to support younger learners in understanding the abstract nature of geometry [55].

Manipulative materials are well-established in early education, particularly in kindergarten, often including pattern blocks, tangram puzzles, Cuisenaire rods, and geoboards [6, 9]. These tools provide young learners with opportunities for tactile engagement and aid in building spatial awareness by exploring geometric relationships [30, 67]. However, traditional manipulatives lack the built-in feedback and scaffolding needed to guide children toward progressively acquiring geometric skills independently [44]. This gap limits children's ability to explore geometric concepts on their own, as they often need teacher assistance to fully grasp these ideas.

With technology increasingly integrated into schools, digital applications that mimic physical manipulatives have become popular. For instance, GeoGebra [67] acts as a virtual manipulative, helping K-6 students understand geometric concepts. However, relying on digital interactions may disconnect children from the hands-on experiences that are crucial for learning [25, 85], since virtual platforms are typically restricted to 2D representations on a screen [49]. Way [82], suggests that combining hands-on activities with abstract tasks enhances mental model development, as physical manipulatives provide sensory engagement that screen-based tools often lack. While traditional manipulatives support embodied learning, they do not offer the interactive feedback available through digital tools [44, 72]. Conversely, digital platforms provide interactivity but may limit the formation of mental models necessary for young learners [1].

To address these limitations, TIEboard is developed to combine the benefits of tactile learning with real-time visual feedback and scaffolding. Unlike TUIs or traditional geoboards, TIEboard leads children through each step of shape creation, helping them grasp geometric concepts with structured prompts. Upon completing each shape, a colored fiber illuminates to signal success, inviting playful exploration through dynamic color variations. This approach creates a solid foundation for structured yet adaptable learning experiences. The next section will examine the theoretical frameworks that informed TIEboard's development, grounding its design in established learning trajectories for early geometric education.

## 2.3 Theoretical Framework: Geometric thinking skills in early age

The primary aim of early mathematics education is to develop mathematical awareness and establish foundational concepts critical for cognitive development [47]. Research suggests that young children have inherent spatial and geometric reasoning skills, which are vital for comprehending both two-dimensional and three-dimensional shapes [8, 10]. Spatial cognition—the ability to visualize and manipulate spatial information—encourages early engagement with geometry [19]. Preschoolers demonstrate their understanding by identifying, constructing, and recognizing geometric shapes, indicating that they can interact with spatial concepts from an early age [12, 66]. Moreover, studies indicate that guided instruction can enhance scientific and engineering skills, especially when paired with hands-on activities, rather than relying solely on exploratory methods [84].

The work of Van Hiele [80] and Piaget [56] highlights the importance of aligning geometric education with children's developmental stages and readiness [12, 27, 79]. The Building Blocks for Little Kids curriculum is a research-based framework that outlines specific learning trajectories for essential geometry topics [12, 66]. This curriculum categorizes geometric learning into key phases: (1) recognizing basic shapes such as rectangles and triangles; (2) shape composition and decomposition; (3) understanding congruence; (4) shape construction; (5) recognizing turns and rotations; (6) measuring; and (7) creating patterns with geometric figures [12, 13]. These phases reflect developmental progression according to Piaget's and Van Hiele's theories, offering a more detailed alternative to broad frameworks [66].

However, many existing manipulatives do not align well with these structured curricula and lack the interactive feedback necessary for independent learning. TIEboard addresses this issue by combining tactile elements with real-time visual feedback and built-in scaffolding, encouraging children's independent exploration of geometric concepts. Inspired by the Building Blocks curriculum [13], TIEboard's modes are specifically designed to assist learning paths such as shape recognition and spatial reasoning. Unlike conventional geoboards that feature static overlays, TIEboard allows children to actively assess and improve their work through interactive feedback, resulting in a more comprehensive learning experience. This structured yet flexible approach makes TIEboard especially appropriate for exploratory learning outside of school, while still being adaptable to classroom environments.

## 3 TIEBOARD

### 3.1 SYSTEM OVERVIEW

The goal of TIEboard is to emphasize physical interactions to create memorable learning experiences. The term *TIE* in TIEboard stands for Tangible Interface for Education, reflecting the focus on tangible and physical interactions. Its form factor is inspired by geoboards, which are frequently used in early geometry education, alongside traditional lacing toys. In order to balance portability and practicality for young learners, traditional geoboards usually measure between 12.7×12.7cm and 17.8×17.8cm [77]. TIEboard's 18×18.5cm size meets these specifications, allowing for easier handling for children as young as 5 years old, and a manageable working area for dynamic visualizations and lacing activities on a standard child-size desk. The 5×6 grid was selected with 7mm holes to balance creativity, usefulness, and ergonomics. This grid size is consistent with lacing toys and geoboards, including such as "The Math Learning Center's" 5x5 grids [78]. Larger grids can make boards difficult for young children to handle, while smaller grids would restrict design options. To keep this balance a customized printed circuit board (PCB) with the above specifications was designed. TIEboard was created to support double-sided functionality, with LEDs purposefully placed next to each hole on both sides of the board. The front side of TIEboard has user-friendly buttons that are easily accessible to children. These buttons serve three distinct functions: the center button lets the user adjust interaction levels, the right button allows to progress to the next step within that interaction, whereas the left button replays the previous instruction in case a child missed an important information. The interface also incorporates a potentiometer to control the glowing colors of laced string. The rear side of TIEboard features five additional LEDs and the Arduino Nano[1]. These extra LEDs at the back allow children to introduce vibrant colors into their laced strings, promoting dynamic visualizations. In order to assure that TIEboard properly promotes geometric learning, many critical design decisions were taken based on insights from related publications in sections 2.1, 2.2, and 2.3:

---

[1]https://docs.arduino.cc/hardware/nano

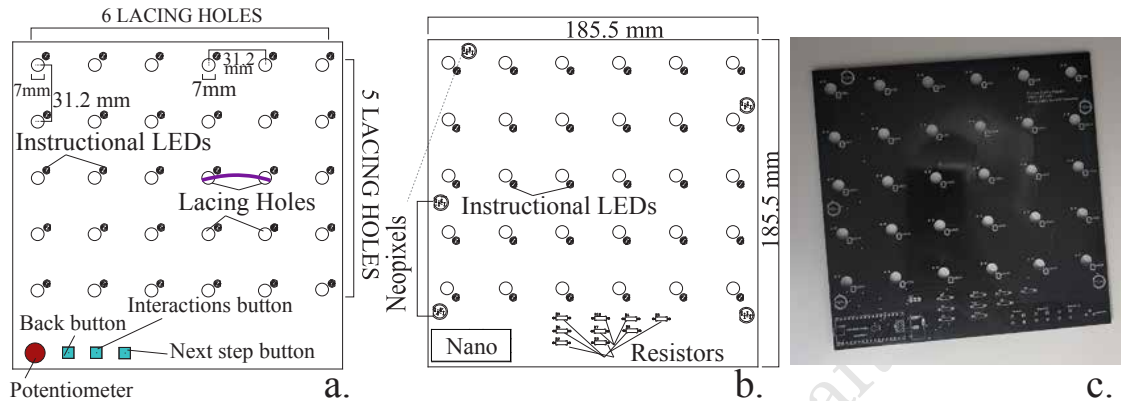

Fig. 3. (a)TIEboard front view , (b) TIEboard back view, (c) Customized PCB created on Eagle software

- Provide scaffolding through step-by-step instructions and interactive activities [20, 63].
- Enhance engagement and sharpen reasoning skills through design activities using a user-friendly interface and built-in instructions that empower users to create intricate shapes not accessible with traditional tools [20, 32, 84].
- Function independently without a computer, making it ideal for flexible educational environments [20, 63].
- Align with the Building Blocks curriculum [13] to help children connect informal and formal learning, enhancing their understanding of geometry and continuity in education.
- Built-in instruction enables autonomous learning while accommodating teacher intervention as needed, all within a size that fits comfortably in the physical space of a classroom[20, 53].
- Promote collaboration, creative learning and is cost effective [20, 41, 42].

## 3.2 DESIGN OF TIEboard

*3.2.1 Instructional Framework.* TIEboard consists of a 5 x 6 grid of 7mm diameter holes, each accompanied by a colored LED to aid in navigation. To utilize TIEboard, users insert a 1.5mm-3mm optical fiber into the holes, guided by instructions on the rear of the board (See Figure 4a). Users power the device by plugging in a battery using the available connection at the back of the board. TIEboard allows the following interactions;

(1) The user first selects the interaction level from the 6 options available (detailed in section 3.3) according to their skills or age by pressing the center interaction button (See Figure 4a).
(2) Once the interaction level is selected, a shape will appear in the form of LED instructional lights. Kids can then start lacing from the back of TIEboard with the optical fiber string inserted in the slot adjacent to the guiding LED (See Figure 4a, 4b).
(3) User then flip the TIEboard and follow the next instructional light to lace accordingly (See Figure 4b).
(4) Once the shape is complete the device will highlight the finalized path by making the optic fiber glow. Kids can then change the color of their shapes dynamically by choosing from the 5 colors provided (See Figure 4d).

The "back step button" (See Figure 3a) allows users to access prior instructions. Although the current system only incorporates the 6-levels presented in section 3.3, TIEboard functionalities can be easily extended to cover more complex concepts via modifications of the ArduinoIDE code, which adds to the flexibility of the tool.

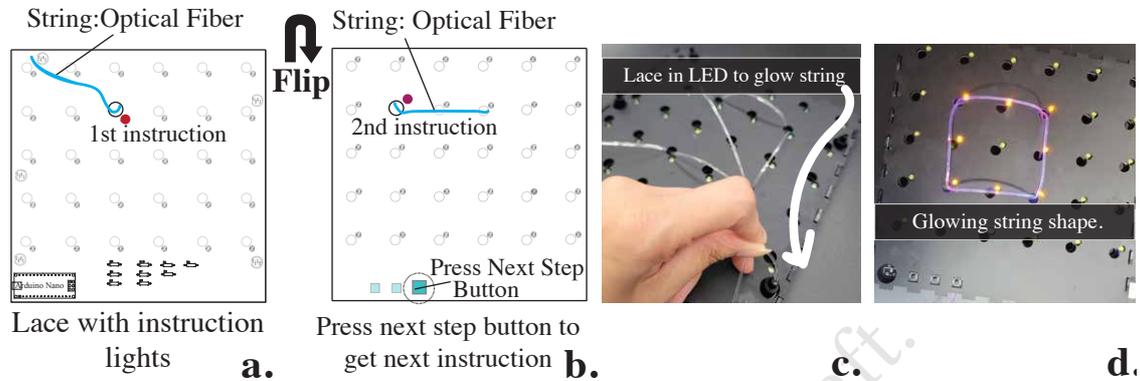

Fig. 4. Instructional Framework: (a) Instructions with LED to start lacing for shape, (b) Press next step button to get next instruction, (c) Insert end of fiber in the LED once the shape is complete, (d) Press next step button to glow the shape with LED

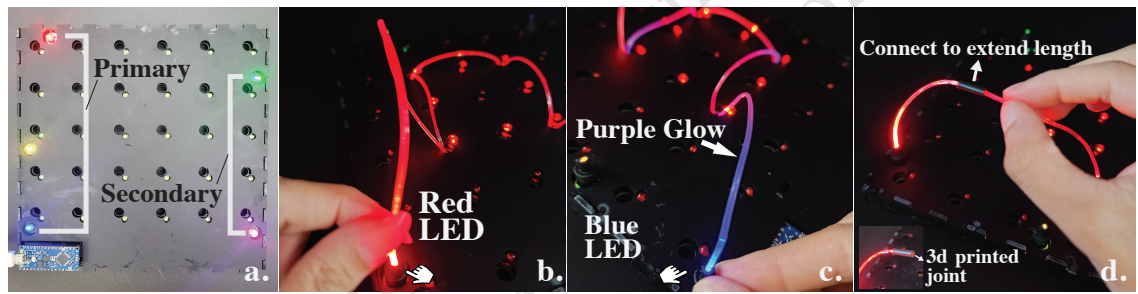

Fig. 5. TIEboard color theory: (a) 3 Primary colors from color wheel (RYB) and 2 secondary colors (b) Insert end of fiber in red LED (c) Mix red and blue color: Insert other end in blue and produce purple color (d) 3D printed part to extend the length of string

*3.2.2 Colored-changing string.* In geoboards and traditional lacing toys, the thread color can not be changed dynamically. To enhance interactivity and playfulness, TIEboard allows children to use optical fibers to create shapes and illuminate their string with customizable colors. The LED system includes five LEDs: three primary colors (red, yellow, blue) and two secondary colors (purple, green), allowing for a diverse range of color combinations (See Figure 5). Users can physically connect the ends of their optical fiber to any of the five LEDs at the back of the board, allowing them to manually create specific color combinations. Additionally, once the optical fiber is placed, children can dynamically change its color by rotating the potentiometer (See Figure 3a), offering further customization without requiring children to move the laced fibre. Moreover, during the Free Play mode(discussed in Section 3.3.6), users can switch between fixed and blinking illumination by pressing the next step button, enabling more creative exploration through pre-coded functionality.

Optical fiber of both 3mm and 1.5mm was utilized to experiment with illumination. While the 3mm fiber resulted in a vibrant glow and a pleasant mixing of colors, its inflexibility prevented it from being laced more than twice through a 7mm hole (See Figure 6a). In contrast, the 1.5mm fiber possessed both a high-quality glow and flexibility, allowing it to be laced multiple times through a single 7mm hole and facilitating the creation of more intricate shapes by children (See Figure 6b). The intensity of the fiber's illumination was enhanced by sanding (See Figure 6c), and the use of a 3D printed part allowed for the extension of the string without compromising its glow (See Figure 5d).

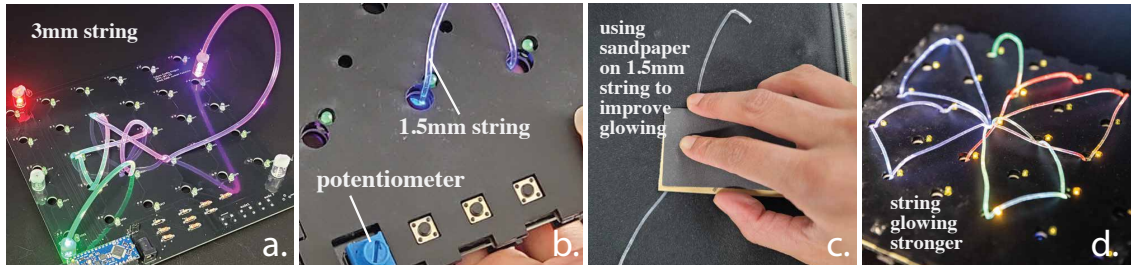

Fig. 6. (a) 3mm optical string can lace only twice from 7mm hole (b) 1.5mm string is flexible, can lace many times from 7mm hole and potentiometer to further control colors of glowing string (c) Sanding the 1.5mm string to enhance glow (d) Enhanced glow after sanding

*3.2.3 Sensing.* One of our design goals was to provide feedback to children during lacing activities without requiring them to stop for the next instruction. As a result, we experimented with a sensing mechanism using stainless thin conductive thread[2] wrapped around the optical fiber and conductive copper tape (resistivity of 0.004 ohm/square[3]) around each 7mm-diameter hole (See Figure 7). An Arduino Nano was used to control the matrix and detect when the string was inserted into the hole, allowing the system to proceed to the next step automatically. Initial testing demonstrated that the system could generally detect lacing activity and provide real-time feedback. However, we also observed how occasional unintended touches resulted in inaccurate detections, which we were concerned could confuse children during activities. As a result, although we believed that preliminary explorations established opportunities for progressively improve automatic sensing accuracy, we decided not to incorporate this function in the device used for the workshops to minimize chances of disruptions for children due to hardware malfunction.

TIEboard's design follows the TUI design principles outlined in Section 2.1, providing a user-friendly interface and structured progression through geometric notions [84]. TIEboard is specifically designed for out of school settings but can be adapted to formal settings. It encourages collaborative learning while also allowing for autonomous or teacher-facilitated use. It improves accessibility across a wide range of educational situations thanks to its low cost (about 25-30 USD per device) and screen-free operation. Furthermore, TIEboard promotes free play by allowing children to explore shapes with light threads inspired by real-world observations or their imagination, so developing reasoning skills through creative learning (See Table 2).

## 3.3 TIEboard Interactions

TIEboard includes interactions with varying levels of difficulty that are tailored to specific age groups, aligning with the theories embraced by the Building Blocks curriculum as mentioned in Section 2.3 [13], and is intended for use by children aged 5 to 9. TIEboard interactions are:

---

[2]https://www.adafruit.com/product/640
[3]https://www.adafruit.com/product/1127

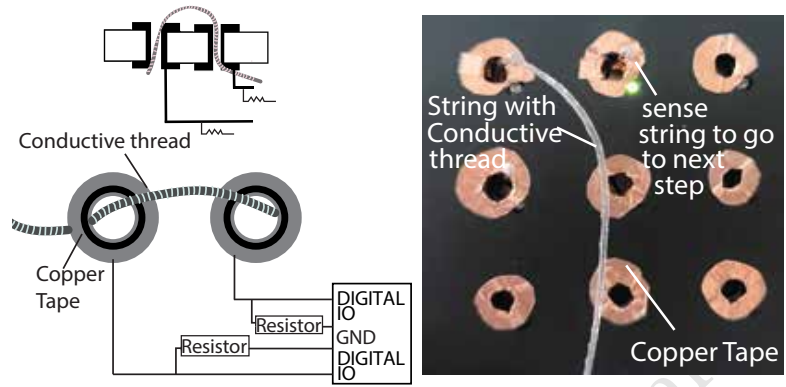

Fig. 7. Sensing circuit with conductive thread on optical fiber and copper tape around holes detects lacing, triggering the next step automatically.

Table 2. Mapping of Design Features to Pedagogical Value

| Design Feature | Pedagogical Value |
| --- | --- |
| Lacing Mechanism | Helps in understanding spatial relationships and geometric shapes with hands on learning. |
| Guided Learning | Step-by-step visual guidance is provided to help with proper lacing patterns, gradually decreasing aid to foster learning independence. |
| Customizable Thread Colors | Encourages creativity and personal expression, making learning more engaging and enjoyable. |
| Multiple Interaction Modes | Adapts to various learning levels and styles, allowing for incremental progress from fundamental forms to complex ones while maintaining continuity with curriculum guidelines. |
| Collaborative Activities | Facilitates teamwork and communication skills through guided collaborative projects. |
| Free Play Mode | Promotes creativity and exploratory learning, allowing children to apply geometric concepts in various ways. |

*3.3.1 Interaction 1: Basic shape.* In basic interaction, users create shapes by following step-by-step instructions, which is influenced by Building Block curriculum's shape identification activity, focusing on understanding the characteristics of basic shapes such as squares, rectangles and triangles. For more complex ones such as hexagon and pentagon, the system features an additional step that assists the learner in creating shapes within shapes to reinforce learning. To start, TIEboard displays the entire outline of the shape with guidance lights, allowing children to become acquainted with the desired end result of the shape they will be lacing. Subsequently, they are provided with step-by-step instructions via LED guidance, which can be automatically dispensed using the sensing function or regulated by the child who can chose to press the next button (refer to Figure 8), when they feel ready to move to the next action. Once the lacing is complete, children can personalize their shapes by illuminating them in an array of vibrant colors.

*3.3.2 Interaction 2: Different orientation and size.* The aim is to teach that shapes are the same even when they're different in size or orientation. This interaction helps to foster an understanding of shape composition and aligns with the principles of the activities highlighted in "Building Blocks" curriculum. This interaction level features complete shape guidance, instead of a one by one instructions, promoting children to engage in the challenge of lacing a full shape on their own, based on previous learning, thus increasing complexity and supporting scaffolding (see Figure 8). After all shapes have been laced, the objective is to identify the odd one. Children are encouraged by the teacher to choose a shape that they identify as different from the others and mark it by making the string glow red (see Figure 8).

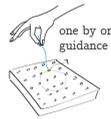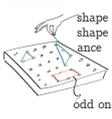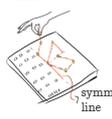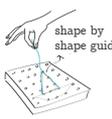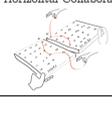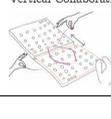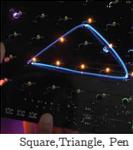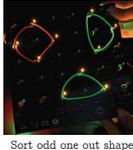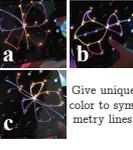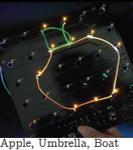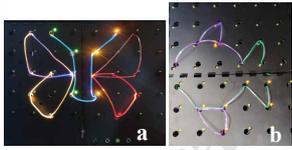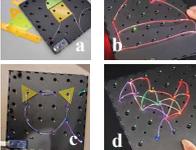

Fig. 8. TIEboard's six different interactions to improve kids geometric learning

*3.3.3 Interaction 3: Symmetry.* In this interaction, children learn about symmetrical shapes with three types of symmetry: horizontal, vertical, and diagonal. Children demonstrate their cognitive ability by memorizing and applying the steps to finish the other half of the shape after lacing only one half of it based on provided instructions. This hands-on approach improves their spatial awareness and allows children to explore congruent shapes using symmetry operations, helping them understand the concept of congruence highlighted in the Building Blocks curricula [13]. To add a creative touch to their geometric exploration, they illuminate their finished creations with dynamic glows and an array of colours (See Figure 8).

*3.3.4 Interaction 4: Complex shape.* The 'Complex Shapes' interaction supports the creation of intricate forms by combining basic shape learned from previous levels, such as showing how a hexagon and triangle could make up an apple. This supports the idea of shape construction mentioned as the 4th component of geometry in Building Blocks curriculum. Additionally, some of the concepts from 'Interaction 2' are integrated, allowing children to create shapes of various orientations and sizes and revisit what they had previously learned. Similar to the "Interaction 2", this level does not provide step-by-step instructions, but features shape by shape guidance, resulting in a more challenging and dynamic learning environment (See Figure 8). Each shape can be personalized using a different color, encouraging creativity and personal expression.

*3.3.5 Interaction 5: Collaboration.* TIEboard's 'Collaboration' interaction aims to promote abstract and both constructive and deconstructive thinking [40, 68]. Participants work together to break down tasks and combine ideas, connecting individual perspectives. The idea of scalability is crucial because it permits collaboration not just between two children but also between multiple children in a horizontal or vertical orientation, akin to a patchwork type of structure (See Figure8), meaning that more children can work together. The system features both guided collaboration via instructional lights, for instance, if two children had to make a butterfly together each child will get half of the instruction on their TIEboard and free collaboration, to enhance kids' engagement. After developing a shape collectively, children can glow their shapes together to create a complex picture which encourages creativity and shared ownership of the results.

*3.3.6 Interaction 6: Free play.* The goal of this interaction level is to enable children to exercise their creativity [14, 42, 57] by exploring patterning to make their own geometric shape based on the concepts learned in prior interactions or according to general guidelines given by the teacher (such as making animals, characters or landscapes with geometric shapes). This aligns with the patterning activity of the Building Blocks curriculum [13]. This interaction offers three different glowing strings through pre-coded options, including fixed color (all red), different colors, and blinking glow, enabling kids to experiment with various lighting combinations by pressing next button (See Figure 4b). TIEboard not only encourages creative exploration but also expressive learning as children actively engage with diverse glowing options, allowing them to express themselves uniquely through the creation of patterns and designs. TIEboard free play can also open up different domains of creative designs like *"Neon Art"* and *"Basic Animation"*. This can also be enhanced by using acrylic modules and other materials in order to increase playfulness (See Figure 8).

## 4 TESTING TIEBOARD

This section presents the methods and results from two workshops conducted with children to assess the impact of using TIEboard for early-age geometric learning. Details concerning participants, structure of the workshops and modalities for data collection and analysis are articulated in the following sections.

### 4.1 Participants

The workshops featured a total of 16 participants, 10 girls and 6 boys between the age of 5 to 9 years old. Our target age group for participation was between 5 and 10 years of age. The workshop participants were distributed as follows: the first workshop featured 12 participants, while second workshop involved 4 participants. Notably, only two children who attended the sessions had prior knowledge and experience with geoboard geometric learning, whereas the remaining 14 were introduced to geometry through teacher-led activities such as whiteboard drawings, cut-out shapes on large sheets of paper, or textbooks. Participants for the first workshop were recruited by their after-school STEAM teacher who helped to organize the workshop. Participants for the second workshop were recruited by contacting a coordinator from the [anonimised] Center, run by a renowned technology company, which organizes various Sunday workshops for children to promote learning. We communicated with partner schools and centers the objectives of the research, the structure of the activity, data collection methods, and the study date and time. A consent form was given to the coordinators one weeks before the study to obtain guardian permission (signature) for their children's participation. The document described the purpose of the study, the nature of participation, confidentiality concerns, and image rights. Each workshop began with the primary researcher outlining the features of TIEboard and guiding participants through its use. This 10-minute session demonstrated how to lace optical fibers and switch between interaction modes. Two facilitators were present in each workshop to support the children by controlling group dynamics, clarifying instructions, and offering individual help with difficult tasks. Parents arrived near the end of the session to see their children's efforts and display their creations. The goals of the workshop were also explained to them, and they were shown visual documentation, such as pictures and the designs the children created during the session.

### 4.2 Procedure

The workshop sessions lasted between 2 to 3 hours and were adapted to the children's pace in tackling the various TIEboard interaction levels. The first author, who was the main contributor to the prototype's development and was well-versed in its functionality, facilitated these sessions. The second author participated in all sessions as an observer, took notes, and provided technical assistance. A Japanese elementary mathematics teacher collaborated with the researchers in the workshops, providing valuable insights and pedagogical support. The children and their accompanying parents were given an explanation of the research's purpose at the beginning of the workshop. After obtaining informed consent and collecting basic information about age and experience using manipulatives such as geoboards, the children were shown how to change modes and elicit instructions from TIEboard. Children were then given time to familiarise themselves with the device before the activities were formally started.

Table 3. Details of TIEboard interactions and participants in the workshops

| | TIEboard interactions | Boys | Girls | Age | Total |
|---|---|---|---|---|---|
| Workshop 1 | Kids engaged with first four TIEboard interactions: (1) Basic Shape (2) Different orientation and size (3) Symmetry (4) Complex shape. Session lasted for about 2.5 hours (See Figure 9). | 4 | 8 | 5-9 | 12 |
| Workshop 2 | Kids engaged with all six TIEboard interactions including (5) collaboration and (6) Free play Session lasted for about 3 hours (See Figure 10). | 2 | 2 | 7-9 | 4 |

Following these introductory steps, geometric exploration began, advancing through the interaction levels of TIEboard's, using the built-in instruction system. The learning pattern progressed from simple to complex interaction level and later experienced incorporating collaboration and free play. The first four interaction levels followed the basic structures outlined in section 3.3. In the second workshop, after completing the first four interaction levels, children were asked to select their partners for the collaboration interaction level. They could collaborate vertically or horizontally, discussing ideas, drawing them on the whiteboard, and creating combined shapes. The final activity involved playing with the free play level for about 30 minutes, during which they could create as many designs as they wanted in the allotted time. As these activities were less structured and did not rely on the embedded guidance of TIEboard, we were conscious of how they could potentially require more dedicated support from facilitators, and relied on more in-depth observation of collaborative dynamics and children's creative processes for analysis. As a result, we opted for limiting the size of the workshop to four children. Although such choice limited generalization due to the small sample, it offered key insights on how collaboration and free play enhance TIEboard's role in fostering creative and cooperative learning. Following completion of the TIEboard activities, children were given the opportunity to demonstrate to their parents how each interaction level worked, explain their collaboration, and exhibit what they created during the free play session. The workshop concluded with the researcher asking the children for feedback on their preferences and dislikes of TIEboard and its interaction levels. They were further asked to provide their opinion about how TIEboard could help their geometric understanding in comparison to current classroom methods. Finally, feedback was sought on potential changes to various tasks, clearer instructions for challenging interactions, and features that promoted enjoyment. During the workshop, it was critical to scaffold activities, ensuring that students finished easier tasks before moving on to more complex ones. The technique supported them to reinforce their comprehension of previous tasks' concepts.

### 4.3 Data Collection and Analysis

The workshop sessions were videotaped and supplemented by detailed field notes taken by the researchers. The first author used interaction analysis [35] to analyze video recordings, specifically seeking to document students' interactions with TIEboard, their interactions with each other and across the classroom, as well as with teachers and facilitators. We analyzed instances first individually and then focused on creating learning trajectories for each child based as they progressed throughout the workshop. Codes for identified interactions were discussed with the other authors, to promote a more holistic interpretation. The use of content logs and annotations from the videos aided in the identification and categorization of recurring themes [11].

For data from the second workshop, we utilize Meier et al.'s [15] nine collaboration dimensions to evaluate the nature of collaboration observed between children while using TIEboard. Each dimension was rated on a 4-point Likert Scale (with -2 denoting 'very bad' and +2 very good') by two independent researchers observing the actions of each of the two groups of children in the second workshop, with each group featuring two participants. This instrument stands out for its ease of use and effectiveness in assessing collaboration, providing a nuanced assessment across multiple dimensions capturing both individual aspects and interactions within the pair, ultimately improving the tool's reliability and usability [15]. Furthermore, we leveraged the Guilford Measures [26], to assess produced outputs during free play along four dimensions, each of which can be evaluated following the principle of divergent thinking, described as the ability to generate multiple ideas following a single prompt [64]. These dimensions include the following: (1) Fluency (number of responses), (2) Flexibility (variety of response types), (3) Originality (unusual or unique responses), and (4) Elaboration (detail and depth in responses). Guilford's measures provide a comprehensive framework for evaluating the number, variety, uniqueness, and depth of creative output and have successfully leveraged to assess children's creativity in STEM activities [7]. Using a similar approach to the one adopted to evaluate collaboration, two researchers utilized the Guilford Measures [26] to independently assess creativity of based on the designs generated by the four children in the second workshop. Independent scores were then reconciled through discussion to reach a finalised agreement.

Across both workshops our analysis was guided by the research questions outlined in the introduction, which focused on children's interactions with TIEboard across various levels mentioned in section 3.3. The research looked at participation during guided tasks, collaboration in group activities, and creativity in free-play settings.

## 5 FINDINGS

The six interaction modes that make up TIEboard are arranged into three primary categories for analysis. The first, "Engaging with TIEboard," discusses the findings from the first four modes, which focused on structured geometric learning. The fifth interaction of TIEboard is examined in the second area, "Collaboration," and the findings from the final free play interaction are highlighted in the third category, "Free Play."

### 5.1 Engaging with TIEboard

This theme mainly addressed RQ1 focusing on understanding children's interactions through the first four levels. From the begining we observed how, step-by-step instructions built into TIEboard, in the first interaction were critical in guiding children's initial experiences with TIEboard. Moreover, the introduction of the triangle as the first shape allowed for a low-barrier entry point with limited geometrical complexity to allow children to focus on getting familiar with the lacing process following the step-by-step instruction. Although initially children were slow and somehow hesitant in lacing the fiber through the holes, their performance improved as they progressed. For example, a 7-year-old child in the second workshop struggled with loose lacing, which affected the clarity of her shapes. Her lacing, however, became tighter as she worked with smaller triangles, which enhanced the shapes' recognition and made it easier for her to evaluate geometric qualities at a later level.

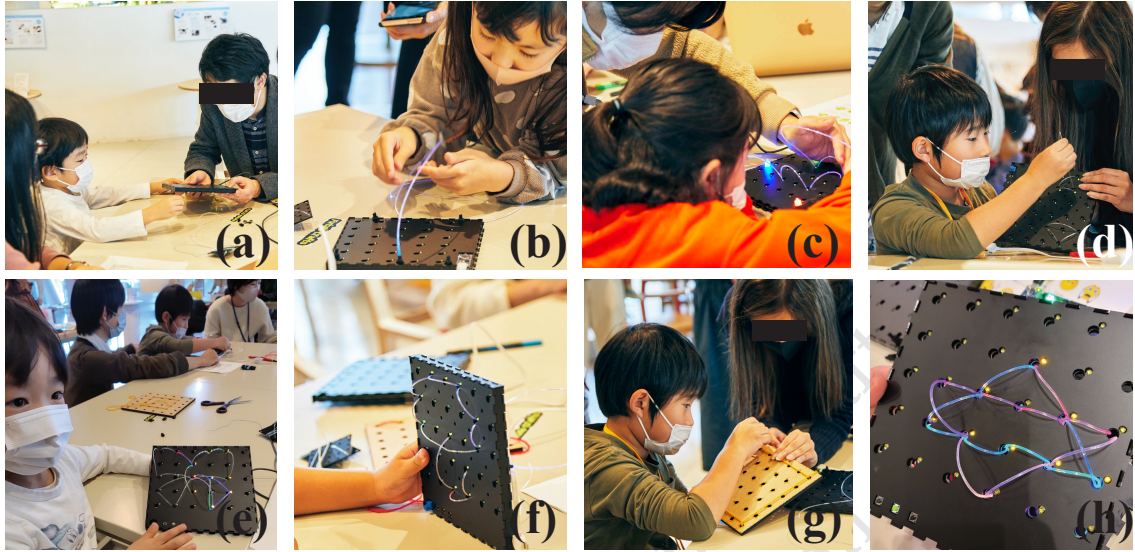

Fig. 9. Workshop session 1: Children engaging with TIEboard's first four interactions

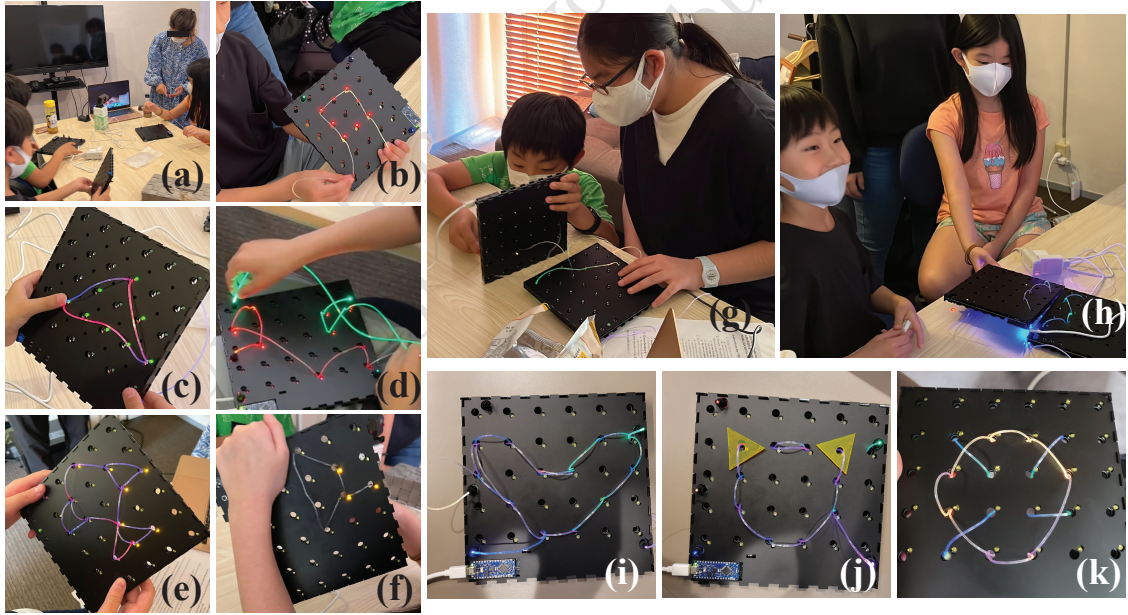

Fig. 10. Workshop session 2: Children engagement with TIEboard's six interactions including collaboration and free play

Children's acquisition of the various components of geometry as articulated in the Building Blocks framework could be observed from very early stages. When creating hexagons in the first interaction, several children were able to draw triangles inside the hexagon without assistance, indicating shape identification. The second interaction required children to identify the odd shape based on their orientation and size. Their reasoning was assessed through observations and verbal feedback. Several children, for example, made explicit observations during the activity. *"Look I found the odd one out shape because it has a different angle than the other two" (Workshop2 - 8 years old boy)*, and *"I recognized the odd one out shape due to different length of sides of each shape" (Workshop2 - 7 years old girl)*. Facilitators also prompted children to describe their choices, demonstrating their ability to differentiate and assess aspects of geometry such as angles and side lengths.

In both the second and third interaction mode, we noticed that the change in the modality of instruction provision created some challenges for children as they expected the board to provide step-by-step instructions, as it did in the previous level. At the start of the activity TIEboard prompts the child to create three triangles, with all three shapes appearing as single instruction, which required lacing and then identifying the odd one out by changing its color. During the workshops, three children in the first workshop and two in the second pushed the 'next step' button, expecting one-on-one instructions, that would cause each of the 9 LEDs to activate in sequence. Instead only the three LEDs of the one odd shape remained lit. Facilitators intervened to explain how the format of the instruction was slightly different for the current interaction level, and to focus on shape-by-shape rather than apex-by-apex prompts. After being explained the change in the instruction modality all children were easily able to successfully complete the activity.

Similarly, the third interaction on TIEboard presented initial challenges as instructions were only provided for one half of the shape, such as a butterfly with a vertical symmetry line. Facilitators encouraged participants who were initially perplexed by receiving only partial instructions to complete the given shapes before attempting to create a symmetrical second half. As they completed one half of each figure, children's confidence increased and the perception of complexity decreased, revealing that the seemingly complicated shapes were primarily made up of simpler ones they were familiar with. This experience helped to make the process clearer and made the initially difficult creations more accessible as the task progressed. Building on the initial understanding, the horizontal symmetry task, in which participants built fish made of two stacked halve, became a smoother task so participants focused on the anticipation of discovering the entire figure and apply their own personal touches to it. As they finished their glowing fish made of triangles and hexagons, the children proceeded to apply various colors to individual strings and proudly displayed their TIEboards to others.

The above observations highlight several important points. Firstly, while children were confused when the instruction format was changed, they were quickly able to follow it after the facilitators clarified it. This indicate that different instruction formats can be used to promote different geometrical skills such as shape composition (Interaction 2), and congruence (Interaction 3), but the TIEboard system should more clearly point out to transition in instructional format, or introduce intermediate steps to facilitate comprehension. On the other hand the progressive scaffolding of the various interaction modes was successful not only in supporting the acquisition of new concepts, but also applying previously learned ones as observed with comments such as *"I can see a triangle in my butterfly (Workshop1- 5 years old girl)"*, ensuring that children were able to decompose more complex figures into simple shapes. During the fourth interaction level where participants created complex shapes to composed objects such as apples, boats, and goggles, no intervention from instructors was required, demonstrating that children were able to follow shape-by-shape instructions. Children also exhibited autonomy in combining shapes to generate everyday objects as well as deconstructing objects into fundamental shapes which shows acquisition of the "Shape construction" principle from the "Building Blocks" framework. We monitored children's completion time across various activities and these data from the sessions revealed a general trend: younger children (aged 5-7) took 7 to 8 minutes on average to complete fundamental forms in Mode 1, but older children (ages 8-9) finished in 5 minutes. In particular, a 9-year-old participant who had previously used a geoboard finished the job in just 4 minutes, demonstrating advanced familiarity and creativity by enhancing things like an apple by dividing it into smaller triangles. Similar trends were observed throughout the following activities, albeit with differences between those with previous geoboard experiences and those without, becoming progressively smaller as the latter group gained more confidence in the lacing motions. These data suggest that younger children need more time and guidance, reflecting their stage of development, while older children displayed better efficiency and independence. However, all children were able to complete activities, regardless of their initial degree of familiarity.

As mentioned in the description of specific episodes the introduction of glowing string, which was encountered for the first time after completing the triangle, added a fascinating dimension to the learning process. All children were excited by the glowing effect's novelty, immediately turning around to show their creations to the others. The shared excitement persisted throughout all workshops and prompted collaborative exploration, with children imitating and enhancing each other's glowing string effects by adjusting the color on their own boards aligning with RQ2. However, throughout the workshops, we witnessed how the use of different glowing colors not only supported engagement and creativity but also facilitated learning. In the symmetry activity, for example, the use of a different color could help to segment the figure, and in the complex shape task congruent and different shapes were often highlighted in different colors to reinforce identification. *"My fish has hexagon and triangle and glowing line of symmetry, with that, I quickly made the other half (Workshop 2 - 9 years old boy)* As workshops progressed, the decreased demand for assistance as well as the ability of children not only to follow instructions, but to continue exploring and modifying their creations, showcased an enhanced grasp of shapes, sides, and angles, with this progress becoming more evident in each successive interaction levels of TIEboard. At the same time, our video analysis revealed that children kept engaging with each other, as well as the teachers, to show their successful creations and exchange ideas about how to illuminate them in different colors. The various features of TIEboard did not just facilitate learning and interaction with the device, but also amongst the people in the classroom. In the following section, we explore more explicitly the findings that emerged from our analysis of collaboration.

## 5.2 Collaboration

The second workshop assessed new modes: "collaboration" and "free play" introduced in TIEboard, after foundational testing with 12 participants in the first workshop. Figure 11 shows the average collaboration scores of TIEboard. Group 1 showed more balanced and effective teamwork than Group 2, as both members actively discussed their plan and drew their collective idea on a whiteboard before starting. A participant with previous geoboard knowledge led the lacing action while helping their partner. The second participant, who enjoyed drawing, enhanced the piece by experimenting with colors, demonstrating active collaboration, as reflected in the following comments. *Child 1 (G1): What hole do you think I should lace now? Child 2 (G1): Lace here to build our shape*. Group 2, on the other hand, started with individual designs and failed to properly define their collaborative approach. This resulted in delays as they attempted to match their shapes for the combination. Despite these obstacles, Child 1 (G2) led the effort to resolve issues, with assistance from Child 2 (G2), the following conversation demonstrates their problem solving approach. *Child 1 (G2): I think we did this step wrong; this string must go from another hole. Child 2 (G2): That's right! Let's take out the string one by one so we do not lose our already laced shape*. Both groups received a score of at least 1 ('good') in six dimensions: individual task management, reciprocal management, technical coordination, task division, information pooling, and dialogue management. Group 2 received a slightly lower score (0.75) in the areas of maintaining mutual understanding and time management.

Moreover, when parents arrived near the end of the workshop, both groups began to display the outcomes and explain how their teamwork unfolded. Group 1 illustrated how they divided the shapes among the members and collaborated successfully to translate their envisioned design onto the whiteboard. Group 2 also recounted their cooperative experience, talking about the obstacles they faced and surmounted through skillful dialogue and negotiation. The clarity with which both groups were able to unpack their processes reflects in-depth comprehension and improved abstract thinking supported by collaborative experience. Finally, when imagining future possibilities to extend the collaboration activities with TIEboard, children highlighted options for combinations which were not limited to pairs. Thanks to TIEboard's square shape, it was mentioned how four children could combine their TIEboards in different directions, forming diverse shapes and illuminating them collaboratively. This shows how, even with relatively limited practice, children could think of creative ways to increase the complexity and variety of the activities available to them.

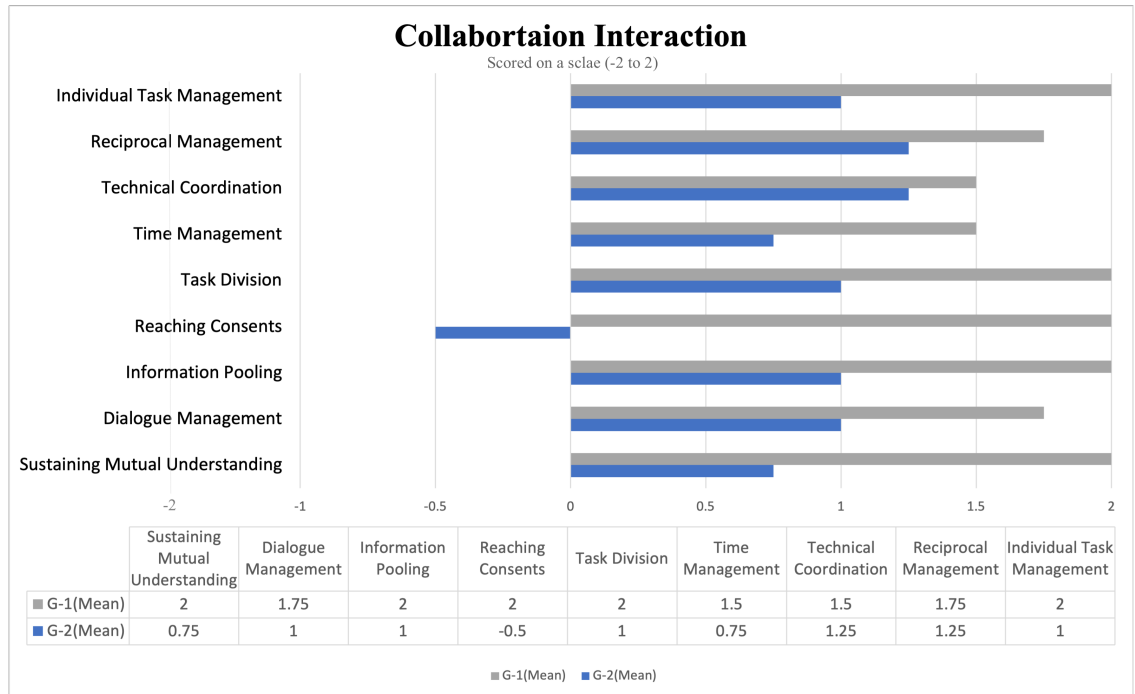

Fig. 11. Collaboration interaction analysis for two groups by two observers, based on the nine dimensions of Meier et al. [15]

## 5.3 Free Play

This section aligns with RQ3 where the instruction given to the children for this activity was to create up to five designs using one or more of three different modalities: single-color light, multi-color light, and blinking light. The researchers evaluated the fluency, flexibility, originality, and elaboration of the children's creative outputs based on both their approach and resulting designs, as shown in Figure 12. Child 1 demonstrated an advanced understanding of multiple geometrical shapes and his ability to build on learned concepts through the creation of two complex designs. These designs not only expanded on the shape-in-shape concept introduced in previous interactions, but his thought process in decomposing the image of a Santa Claus and a Christmas ornament in the first composition and a crown in the second composition into recognizable objects highlighted a high degree of thoughtfulness and detail in his designs. As a result, researchers attributed to his work the highest elaboration score. Child 2 demonstrated exceptional fluency, creating designs inspired by her interest in kittens and cats. While her designs featured recurring motifs like triangles and octagons, her ability to iterate multiple variations efficiently highlighted her creative speed and assurance. Child 3 showed outstanding originality, producing three designs that creatively disrupted conventional geometric rules. For instance, his use of symmetry in two designs and composite shapes in another reflected a willingness to experiment and innovate. This originality illustrates how TIEboard encouraged him to think beyond basic geometric patterns and explore novel combinations. Finally, Child 4 showed the most flexibility by producing a wide variety of response types, each more prominently inspired by a specific geometric concept such as lines, angles or interlocking shapes.

|  | Fixed Light | Different Light | Blinking Light | Design 1 | Design 2 |  |
|---|---|---|---|---|---|---|
| Child 1 Boy 8 years | 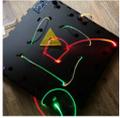 | 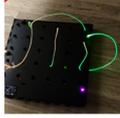 |  |  |  | Highest Elaboration |
| Child 2 Girl 7 years | 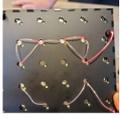 | 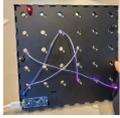 | 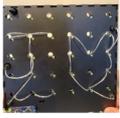 | 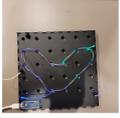 | 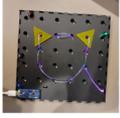 | Highest Fluency |
| Child 3 Boy 7 years | 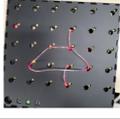 | 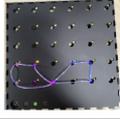 | 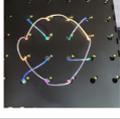 |  |  | Highest Originality |
| Child 4 Girl 9 years | 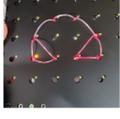 | 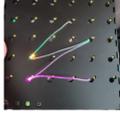 | 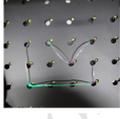 | 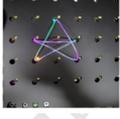 |  | Highest Flexibility |

Fig. 12. Free play analysis based on Guilford measure [26]

The lighting options for the fiber optic strings, especially the blinking ones, inspired children's actions. The feature sparked Child 1's Christmas design ideas. Child 2 said the light made her kitten design cuter. The blinking light version was Child 3's favorite; he used it to design a moving wheel. Finally, Child 4 explained that the blinking of the fiber reminded him of the stars which motivated him to create his final design. The evaluation of the Free Play session using Guilford's metrics provided a structured yet flexible framework to analyze the children's creative processes. While the assessment relied on qualitative observations rather than numerical scoring, this approach was purposefully chosen to align with TIEboard's educational goals of fostering creativity and enabling open-ended exploration. By focusing on fluency, flexibility, originality, and elaboration, the evaluation captured the diverse ways in which children engaged with geometric concepts and expressed their ideas. This method emphasized observing how children applied their prior learning, generated unique designs, and adapted to new challenges presented by TIEboard. While acknowledging the subjectivity inherent in qualitative assessments, this approach ensured a holistic understanding of how children interacted with the tool.

## 6 DISCUSSION

The results of our workshops, show how TIEboard complements curriculum-based ideas in geometry education, enhancing the usefulness of TUIs for children and contributing to the current discussion on striking a balance between guided and exploratory learning with appropriate scaffolding[24, 43, 51, 84]. The instructional framework which incorporates instant feedback provides an initial support that facilitates learning of basic geometrical concepts. At the same time, the modular system of the board, which allows for the creation of complex structures through collaboration, supports progressively more sophisticated learning. Finally, the Free Play mode and light customization promote creative exploration, in line with Resnick and Silverman's *low floors, high ceilings and wide walls* concepts [62]. Although these concepts have traditionally been used in TUI design for computing [4, 63, 75], they are also useful in geometry teaching. Our workshops clarified areas where TIEboard's design may be improved by lowering floors, raising ceilings, and widening walls ensuring that children can be better supported in their learning journey. In the following sections we discuss our insights on how TIEboard and future digital manipulatives for geometry education can better incorporate two key aspects (1) Facilitating exploratory learning and scaffolding progress, and (2) Enabling collaborative and expressive learning.

### 6.1 Facilitating exploratory learning and scaffolding progress

The outcomes of both workshops showed that children were able to interact with TIEboard and complete the first four levels of guided activities with minimal intervention from the facilitators. Although only two children had previous experience with geoboards, the process of lacing the optical fiber through the loops to create various shape proved to be intuitive for all participants. Children, including those who initially felt hesitant, reported increased ease and engagement with the activity over time, which was further confirmed but their increased speed in creating shapes. As geometry learning is deeply connected to spatial and embodied cognition, ensuring that the gestures and interactions for digital manipulatives are familiar to the child and reinforce logical links with the concepts to be learned was a crucial factor in the success of TIEboard [71, 76]. The built-in instructions offered critical scaffolding, promoting early geometric learning as highlighted in studies on structured guidance [27, 84]. Immediate feedback through TIEboard's glowing fiber provided both motivation and reinforcement, aligning with prior research on integrated feedback mechanisms in TUIs [44, 84, 87]. The findings underline the importance of coupling instructions with tangible rewards to create a more engaging learning experience[45, 56, 76].

Previous studies have shown physical interaction and conceptual learning are connected, meaning that increasing the complexity of one could also affect the other [1]. Throughout the workshop, TIEboard's incremental progression from simpler shapes to more difficult interactions significantly enhanced participant enjoyment and facilitated learning. Our design, drew inspiration from established frameworks, such as the Building Blocks curriculum [13], to build a progression for geometry learning. Researchers in digital manipulatives for STEAM emphasize the importance of integrating scaffolding with established curricula to enhance comprehension of abstract concepts [17, 56, 63, 80].

### 6.2 Enabling Collaborative and Expressive Learning

The idea of *widening walls* to allow children to express themselves and engage in learning activities in a way that is meaningful to them, has arguably been described as the most important feature that should be accounted for in the design of digital manipulatives [40, 62, 63]. The results of the workshops showed how the various features of TIEboard supported creative exploration and personalized expression in a number of way. Firstly, the ability to be able to illuminate one's compositions in one's preferred manner using different colors and blinking patterns led to playful moments. This became even more evident during the Free Play session where the blinking light glow sparked ideas for creating festive-themed shapes, from Christmas-themed decorations to sparkling stars, demonstrating TIEboard's dynamic and stimulating nature in encouraging creative endeavors.

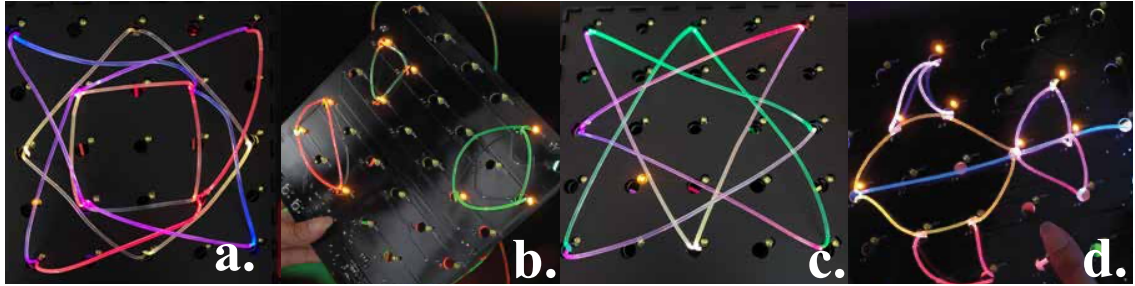

Fig. 13. (a) Reinforce understanding of geometry by creating shapes within shapes, (b) Learning odd one out shape with unique color,(c) Make intricate geometric shapes, (d) Teaching symmetry like horizontal symmetry with fish

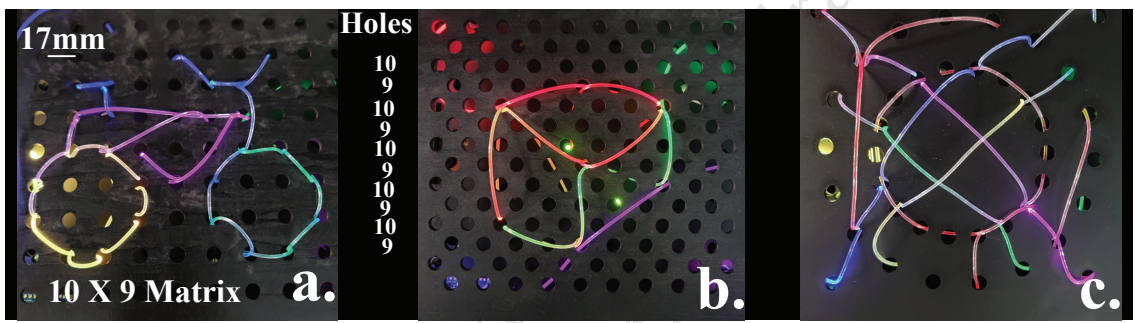

Fig. 14. (a) TIEboard holes with 10X9 matrix with 17mm spacing (b) Isometric TIEboard template with varying holes in each row, (c) Circular TIEboard template

Moreover, the Collaboration mode allowed children to explore the possibilities of combining their efforts to create new and different designs. The result presented in section 5.2 illustrated how the collaboration process could at time be challenging for children to manage, suggesting that providing some initial framing might be beneficial to explore the advantages and disadvantages of different approaches [40]. The form factor and interaction capabilities of TIEboard also offers significant opportunities to expand collaboration, going beyond pairs or small groups in which TUIs are generally evaluated [4, 40, 63]. It is possible for an entire classes to combine their TIEboards in a patchwork fashion, creating a variety of glowing shapes together. Creating complex patterns with friends, does not just represent a way for geometry focused tangible manipulatives such as TIEboard to expand conceptual learning, but also to promote inclusive and dynamic learning environments by facilitating larger-scale collaborative engagement (See Figure 15a).

TIEboard offers an opportunity to widen walls and support creativity by connecting geometric learning to other activities such as computing. The interaction with the Arduino board enables the creation of customized patterns through various programming interfaces, as depicted in Figure (15a). As an example, TIEboard various glowing effects allow kids to make basic animation [4] via blinking LEDs by setting the delay time for each LED to create the moving illusion of the shapes (See Figure17c). In addition, by incorporating additional holes in the outer frame, TIEboard allows users to experiment with shapes by threading and looping the thread around the 3D-printed pins. This combination of holes and pins facilitates the creation of intricate and organic shapes, reminiscent of unique neon art, as illustrated in Figure (16c and d). In future iterations of TIEboard, as well as in the development of future TUIs we believe that taking advantage of both physical and digital features offers unique possibilities for children to truly unleash their creativity. The adaptability of digital manipulatives can improve STEAM activities in subjects such as geometry, computing, and arts, indicating greater implications for learning methods, though current research focuses on digital tools appropriate for older students[34, 58]. After systematically investigating the impact of TIEboard on geometric learning, improved collaboration and creativity, it is evident that this creative tool differentiates itself by supporting a highly engaging learning environment.

---
[4]https://www.dragonframe.com/introduction-stop-motion-animation/

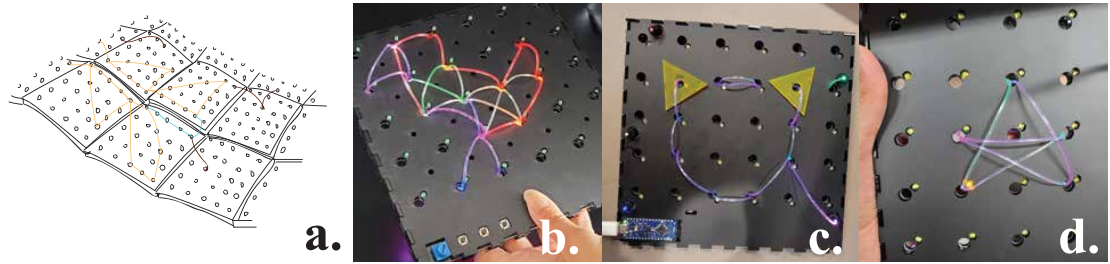

Fig. 15. (a) shows the endless collaboration possibilities with TIEboard (b), (c) and (d) shows creative art with the glowing string (b) shows duck shape created by making shape in shape, (c) acrylic modules are used to create a cat and (d) shows star creation

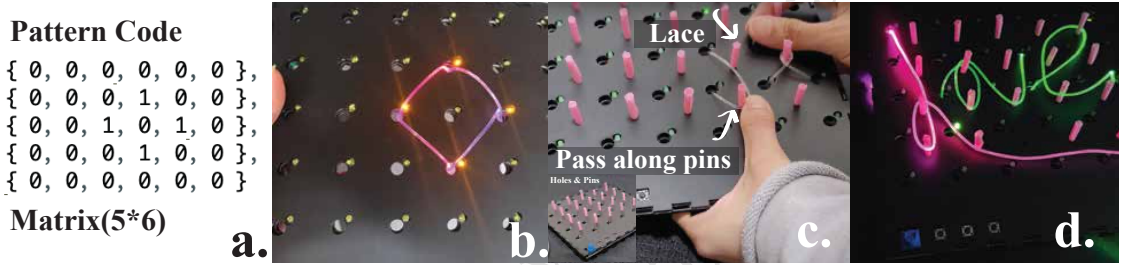

Fig. 16. (a) Code for creating a custom pattern on Arduino IDE using a 5x6 matrix (b) Lace your coded pattern (c)Add 3D printed pins according to your design (d) Lace through the holes and pass along the pins to make organic designs

## 7 LIMITATIONS AND FUTURE WORK

While TIEboard demonstrated strong engagement and learning potential, several areas for improvement emerged through workshop observations and participant feedback. Children's confusion during the second interaction level, which featured a shift in instruction style, highlights how care should be taken when altering the format of instructions to suit progressive learning goals. This aligns with research emphasizing the importance of scaffolding during transitions between learning stages, particularly in geometry education [56, 80]. The confusion was not only due to changes in the instructional format but also related to the confidence children needed in their lacing skills. To address this, we recommend progressively guiding and scaffolding students into more challenging tasks to help them gain confidence and promote developmental learning [62, 80, 84]. Maintaining the presence of step-by-step instructions, where the fiber path is gradually revealed, would help lower the "floor" for learners who need additional support. Conversely, a selectable mode offering shape-by-shape prompts may benefit more confident learners. These insights highlight opportunities for refinement, such as offering customizable guidance modes to better support diverse learning styles. By tailoring scaffolding to individual confidence levels, TIEboard could lower its "floor" for beginners while maintaining its "ceiling" for advanced users, as emphasized in prior TUI research [4]. Our experience also suggests that decoupling physical complexity from conceptual complexity could further support learner autonomy and pacing [1].

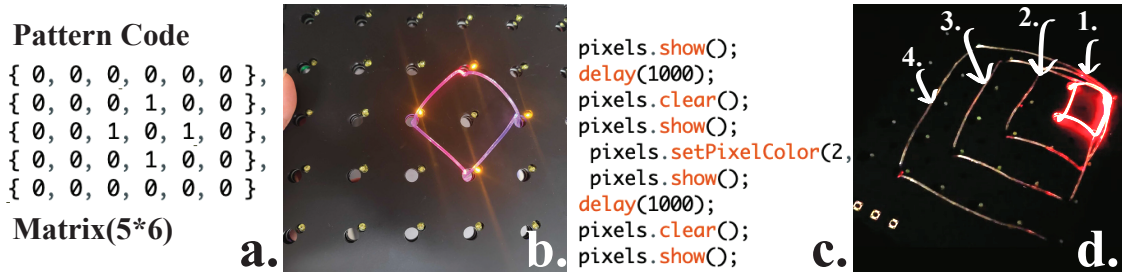

Fig. 17. (a) (b) and (c) shows how to set delay time for LEDs to create your own basic stop motion animation (d) fiber labeled (1) will glow first then 2 followed by 3 and 4)

Although the existing TIEboard levels help children progress through a series of increasingly sophisticated geometrical concepts, the system's overall complexity remains physically constrained. Challenges in "raising the ceiling" of TUIs are well documented and often stem from their tangible nature [1, 4, 17, 28]. However, TIEboard's modular nature provides insights into overcoming these constraints. While the size and shape of the board are fixed, the number and arrangement of holes can be easily adjusted through interchangeable templates. We developed additional configurations, including larger 10×9 matrices, circular grids, and isometric patterns (see Figure 14), enabling more intricate and varied geometric constructions. The sensing method embedded in TIEboard, which was experimentally tested in the lab (as described in Section 3.2.3), demonstrates potential for integration with digital systems. Although not implemented in the workshop due to concerns about misdetection that could disrupt learning experiences, this approach could pave the way for hybrid systems that expand the complexity of TUIs, facilitating seamless exchange of geometric patterns and collaborative learning [4, 28].

## 8 CONCLUSION

This paper proposed TIEboard, a new digital manipulative tool which supports geometry learning in early education. Holding a string of optical fiber in their hands and lacing it around the TIEboard to learn to create shapes, following a progression of activities based on established curricula, provides children with multiple sensory experiences that improve their learning experiences and promote engagement. The TIEboard's built-in instruction system allows users to create complex designs and learn in real time with little reliance on external help; however, initial scaffolding and occasional instructor intervention helped children bring their ideas to life and maximize engagement with the various interaction modes. The system also promotes collaborative activities that can be carried our between pairs of children or potentially extended to a whole classroom. The flexible combination of computational abilities and glowing strings can be leveraged to introduce various interaction techniques with applications ranging from embodied geometric learning, creative art, animation, and tangible gaming for kids. Our study has shown how TIEboard can represent an effective and engaging tool for children between 5 and 9 years of age, as well as being easily adaptable to individual preferences, and potentially supporting extensibility for trans-disciplinary activities. TIEboard's future developments could include enhanced error detection and immediate auditory feedback to let users self-correct with minimal guidance. Furthermore, adding Wi-Fi connectivity would allow TIEboards to communicate, creating the potential for collaborative shape-building games and engaging activities with several players.

## ACKNOWLEDGEMENTS

This study was supported by the Japan Society for the Promotion of Science (JSPS) through the Early Career Scientist Grant-in-Aid [Grant Number: 22K17939].